\documentclass{eas}
\usepackage{graphicx}
\usepackage{amssymb}
\begin{document}

%%-----------------------------
%%      the top matter
%%-----------------------------
\title{The Origin of early $^6$Li and the Reionization of the Universe}

% Short title
\runningtitle{Hubert Reeves : The Origin of $^6$Li
\dots}
\author{Hubert Reeves}
\address{Centre National de la Recherche Scientifique, Paris }
%\secondaddress{his/her second address}
%
%\author{author two}
%\address{his/her address}
%
%\author{author three}
%\address{his/her address} OR \sameaddress{1}
%
%\thanks{thanks}
%
\begin{abstract}
Observational data on the early galactic abundances of the 
light elements lithium, 
beryllium and boron are combined with data related to the reionization of the 
intergalactic medium (IGM) in a search of processes happening 
in the early universe. 
Early massive metal-free stars ( Pop III), largely held responsible for the 
reionization of the IGM, are proposed to have been also the injectors of the 
lithium-6 plateau, through their winds. In this sense the evolution 
of the $^6$Li/$^9$Be ratio appears to be a key parameter for the history 
of nucleosynthesis as a monitor of the early formation of metals and their 
subsequent injection in high energy particles. 

\end{abstract}
\maketitle
%%-----------------------------
%%      your text
%%-----------------------------
\section{Introduction}
In this paper, three groups of ``unexpected and surprising observations" 
are considered in an effort 
to obtain information on the physical processes happening in the 
distant past of the cosmos. 
We consider first the detection of $^6$Li in low-metallicity stars 
at a level far 
above the value predicted by Big Bang Nucleosynthesis (at least a 
thousand times larger). 
The so-called $^6$Li plateau (analogous to the $^7$Li Spite plateau) 
extends from metallicity 
of -2.7 to -1 in logarithmic solar units (Asplund et al 2003, Bonifaccio 2002, 
Cayrel et al 1999, 
Hobbs and Thorburn 1997). Galactic Cosmic Ray (GCR) production by 
collisions of alphas 
on alphas appears to be the most likely formation mechanism. But 
in the framework of the 
presently accepted picture of stellar evolution and nucleosynthesis, 
the presence of $^6$Li, 
at the measured value and at the lowest metallicities poses a major 
energetic problem. 
The question is: what mechanism can be held responsible for that 
much $^6$Li accompanied with so little metals ? 

Second, we consider the fact that the stellar abundances of beryllium and boron increase 
linearly with the abundance of Fe (Garcia-Lopez 1999, Primas 1999) or that, in other words, 
the ratios of Be/Fe and B/Fe remain constant with metallicity. These observations imply 
that the Be and B were ``primary" elements in these early times 
and not ``secondary" as they are 
mostly today (Duncan et al. 1992 and 1997, Prantzos et al. 1993, Cass\' et al. 1995).
How can we explain the constancy of the Be/Fe ratio? A correlated question is: 
why is the $^9$Be/$^6$Li ratio increasing with metallicity (since they are ``naively'' thought
to be both produced as primaries) ?
 
Third, we consider recent data on the reionization of the IGM 
at the end of the dark ages (Madau and Rees 2000). Data in the QSO lines of sight, interpreted 
in the framework of the Gun-Peterson effect, show that the reionization was largely completed 
at redshift  z = 6 ( less than $1\%$ neutral matter remaining). 
Studies of the fossil radiation (CMB) by the WMAP satellite (Cyburt 2004) go back even further. 
The surprisingly high opacity observed ($t\simeq 0.17$), if confirmed, implies that the reionization 
may have been half completed at z = 10 or more, at a time when metal abundances were presumably 
extrememely low. The question is then: what mechanism could have generated the amount of energy 
needed for ionization while forming so little metals ?  
It will be interesting to note that the energetic requirements for $^6$Li formation are 
quite similar to the requirement for reionization thereby offering a possible solution 
to the energetic problem (Lambert 2004) attached to their formation.

\section{$^6$Li observations and energy requirements}

The $^6$Li/H plateau has a value of
$^6$Li/H$\simeq$ 10$^{-11}$ between metallicity -2.7 and -1.
 However, since the WMAP observations (Cyburt 2004) 
have shown that the $^7$Li plateau (Spite and Spite 1982, Chen et al. 2001) has been depleted by a 
factor of around three by stellar atmospheric processes, and since $^6$Li is more fragile than $^7$Li, 
the surface depletion of $^6$Li should be at least as much. In the present interpretation of the $^7$Li 
depletion by atomic diffusion and turbulence (Richard et al 2005, Th\'eado \& Vauclair 2003, 
Lambert 2004, Charbonnel \& Primas 2005), the same depletion should apply to both isotopes. 
Hence we adopt a plateau value of $^6$Li/H$\simeq 3.10^{-11}$. 
The energy required to generate the light element production by high energy collisions 
has been computed by several authors: (Ramaty  2000, Prantzos  2005). 
The cross-sections for Li production by alpha+alpha reactions is substantial (tens of millibarns) 
only in the range from 40 to 200 MeV (Mercer et al 2001). At higher energies the cross-sections 
are negligible. Computations show that the generation of the observed $^6$Li value by fast alphas 
in this energy range requires around 300 eV/nucleon of the ISM  gas. On the other hand, computations 
of the stellar nuclear energy released at 
[Fe/H]=-2.7 (metallicity of Z = 2 10$^{-5}$) yield a value of 
$\simeq 160$ eV/N (8 MeV/N for each metal). 
Clearly an unreasonable efficiency of conversion of stellar energy in fast particles 
would be required to account for the $^6$Li formation. 
The same conclusion has been reached by Prantzos(2005)
from an analysis of the supernovae yields of Fe. Observations of $^6$Li in even lower metallicity 
stars would reinforce this conclusion.

Thus we must look for a process that generates $^6$Li accompanied by very little nucleosynthesis. 
Ideas have been presented by Jedamzik (2000, 2004) (exotic  unstable particles 
with  appropriate lifetimes) 
and Suzuki and Inoue (2002) (gravitational collapses of early structures in combination with 
magnetic acceleration). See Prantzos (2005) for a critical discussion.

\section{Linearity of Be/H versus Fe/H}

To discuss the linearity of the Be/H vs Fe/H we consider first the present day mechanisms of 
formation of Li-Be-B, namely GCR impinging on the ISM.  
Two different processes are simultaneously at work:
1) Fast p and alpha on CNO in the interstellar medium (ISM); the formation is called ``secondary" since 
the ISM CNO was generated by earlier stellar generations. 
2) Fast CNO on p and alpha in the ISM; the formation should primary as suggested by Be and B 
observations,
implying that GCR had always the same CNO content (Duncan et al. 1992).
Today, 
the process 1 represents 80 \% of the yields of LiBeB and process 2 only $20\%$ 
(Meneguzzi, Audouze and Reeves 1971) . Since 
the abundance of CNO in the ISM  was smaller in the past, this ratio must have been different 
(higher fractional contribution of process 2) .

An important information on this subject comes from the 
observation of the ``abnormal"  $^{22}$Ne/$^{20}$Ne 
ratio in GCR today. The fact that it differs notably from its value in the local cosmic abbundances 
is usually interpreted by assuming that a fraction (perhaps as large as 10 to $20\%$, Meyer 2005) 
of the fast CNO (process 2) were injected by strong stellar winds, usually associated 
with emission from  Wolf-Rayet stars (WR).  We shall refer to this process as the 
``stellar wind component of GCR". The origin of the remaining fraction of process 2 has been 
a matter of debate.  One  often held view  is that it represents injection from supernovae ejecta.
However the absence of radioactive $^{59}$Ni (lifetime $\simeq$ 10$^5$ yr)
in the GCR is usually interpreted as imposing a 
delay of several hundred thousand years between the supernova (SN) explosion and the a
cceleration of particles by magnetic shocks; this  implies a dilution of the ejecta in the 
ISM and hence an important decrease of the accelerated  ratio of CNO/H. 
Models of superbubbles have been invoked to circumvent this difficulty (Parizeau and Drury 1999, 
Ramaty et al. 2000), but they remain controversial (see Prantzos 2005 for a discussion). 
Without entering in this debate, we note that the stellar wind component of process 2 could,
by itself, account for the primary character of the early Be and B formation if this process 
was actually the main source of LiBeB in the early years of the galaxy (Cass\'e et al. 1995). 

One difficulty with this hypothesis comes from the assumed presence of WR stars in the early 
life of the galaxy. Since surface mass ejections and stellar winds are associated with atmospheric   
opacities and hence with the presence of metals, it is expected that low metallicity stars 
should not have strong stellar winds. However surface mass ejection could occur also from other 
mechanisms. As an example, Maeder and Meynet (2005) have recently computed models of 
rotating early stars whose equatorial zone rotation reaches the break-up point, 
leading to large mass ejection.

\section{The energetics of the reionization of the IGM}

Photons of at least 13.6 eV are required to ionize H atoms. 
However, because of later recombination, 
it is usually considered that several ionizing photons per H atom 
are required to maintain the ionization. 
The energy expenditures should thus be of the order of 100  eV/N. We note that this amount 
is of the same order of magnitude as the amount required for $^6$Li formation.  
Only stars with 
masses above 3 solar masses emit a substantial amount of ionizing photons. The number of ionizing 
photons per generated atom of metals, for  a Salpeter Initial Mass Function (IMF)
is of the order of 3 10$^4$ . With this IMF, complete 
reionization would correspond to a metallicity Z of 3 10$^{-4}$ (about 0.03 solar). 
But in the absorbing 
clouds observed in the QSO lines of sight, this value of Z is reached at redshifts of z=3, 
way after the reionization period... 
Presumably the early IMF was tilted toward more massive stars. However even the maximum number 
of ionizing photons per metal (8MeV/N/100eV/N) corresponds only to  a metallicity
Z=$10^{-5}$, still most likely 
far too high to have been reached at the reionization period. The question is then: what mechanisms 
could have reionize the universe while generating so little metals ? 

Studies of high z galaxies by  Lehnert and Bremer (2004)  have lead these authors to the conclusion  
that although quasars play an important role in keeping the IGM ionized today, this was most 
likely not the case in the early universe (too few quasars at that time, denser universe, hence 
shorter recombination times). Their observations of early galaxies lead them to conclude that 
the first generations of stars (so-called  POPIII) are the most likely sources 
of the reionization of the IGM . 

\section{Conclusion and questions}

Putting together the analysis of the three groups of observations reported in this paper,  
a tentative coherent scenario is suggested. Stellar atmospheres of early hot metal-free stars (PopIII) 
(Weiss 2000, Wyithe  and Loeb 2004), responsible for the reionization of the IGM, could also be 
the injectors of the early GCR. In the early metal-free universe these GCR would have generated 
the $^6$Li by the alpha-alpha reactions. Later, as the CNO abundances increased in hot stars, 
the GCR started to produce Be and B, thus explaining both the decreasing $^6$Li/Be ratio and 
the primary character of the Be/H and B/H vs Fe/ H. In this sense the evolution of the $^6$Li/Be 
ratio appears to be a key parameter of the history of nucleosynthesis as a monitor of the early 
formation of metals and their subsequent injection in high energy particles.

One question remains: why did those early stars generate so little metals 
(Iwamoto et al 2005)? Remembering that very massive stars are expected to end up 
largely as black holes, or to have experienced during the explosion process a ``mass cut" different 
from the supernovae of today , is it possible that these early stars would 
have mostly retained their metals inside?

%%-----------------------------
%% BIBTEX use
%%-----------------------------
%% To use BIBTEX uncomment the two '\biblio..' lines
%% ATTENTION: don't forget to add the content of the .bbl 
%% file in your final .tex file before submission and comment
%% the next two '\biblio...' lines
%
%\bibliography{apj-jour, add your bibliography file}
%\bibliographystyle{astron}

%% If you are using BIBTEX replace the following lines
%% by the content of the .bbl file
%% 
%%-----------------------------
%%      your bibliography
%%-----------------------------

\end{document}